# Effect of oxygen incorporation on normal and superconducting properties of MgB$_2$ films

R.K. Singh,[1] Y. Shen,[1] R. Gandikota,[1] C. Carvalho,[2] J.M. Rowell,[1] and N. Newman[1,a)]

[1]School of Materials, Arizona State University, Tempe, AZ 85287-8706

[2] Departamento de Física e Química, Universidade Estadual Paulista–UNESP, Brasil

Oxygen was systematically incorporated in MBE grown MgB$_2$ films using in-situ post-growth anneals in an oxygen environment. Connectivity analysis in combination with measurements of the critical temperature and resistivity indicate that oxygen is distributed both within and between the grains. High values of critical current densities in field (~4x10$^5$ A/cm$^2$ at 8 T and 4.2 K), extrapolated critical fields (>45 T) and slopes of critical field versus temperature (1.4 T/K) are observed. Our results suggest that low growth temperatures (300$^o$C) and oxygen doping ($\geq$ 0.65%) can produce MgB$_2$ with high J$_c$ values in field and H$_{c2}$ for high-field magnet applications.

Since the discovery of superconducting MgB$_2$,[1] there has been a worldwide effort to understand the properties of this two band superconductor, and to develop it for use in commercial applications such as high-field magnets.[2]

In clean MgB$_2$ wires, H$_{c2}$(0) has been found to be only 16 T.[3,4] A large increase in $H_{c2}$ of MgB$_2$ in bulk has been shown to result from alloying with carbon,[5-9] and also, in the case of thin films, from disorder and impurities introduced during growth and from irradiation by energetic particles.[9-11] Improvements in J$_c$ have been found in MgB$_2$ thin films grown in an oxygen atmosphere[12,13] and in bulk samples when alloyed with carbon,[14] SiC,[15,16] SiO$_2$ and silicides of W, Zr and Mg.[16]

With magnesium, and to a lesser extent boron, being readily reactive with oxygen to form their respective oxides, significant level of oxygen are typically incorporated into MgB$_2$ during synthesis.[17] These contaminants affect the superconducting properties. Oxygen is incorporated in bulk MgB$_2$ in various forms: as nanometer-sized coherent MgB$_{2-x}$O$_x$ precipitates in the grains,[18,19] as MgO particles in the interior of the grains,[20-23] or between the grains as thin (~3 nm) BO$_x$-MgO$_y$-BO$_z$ layers[24] or thicker MgO layers.[25,26] While the coherent non-stoichiometric precipitates and MgO particles in the MgB$_2$ matrix improved flux pinning resulting in higher $J_c$,[24,27,28] the MgO layers at the grain boundaries reduced $J_c$, presumably as a result of the diminished connectivity.[25] Oxygen rich $MgB_2$ thin films are reported to contain MgO particles[29] and exhibit enhanced $H_{c2}$ and $J_c$ values.[29,30] In all these reports, however, the effects of oxygen content in $MgB_2$ have not been systematically studied.

We report here the dependence of $T_c$, resistivity, connectivity, $H_{c2}$, and $J_c$ of $MgB_2$ films on the oxygen content of the films. $MgB_2$ films were deposited from a Mg Knudsen cell and a B electron-beam evaporation source in an all metal seal MBE chamber with unbaked base pressure of ~$10^{-8}$ Torr and growth pressure of ~$2 \times 10^{-7}$ Torr on c-sapphire substrates at 300±2 °C, as described in more detail elsewhere.[31,32] Oxygen was incorporated into the $MgB_2$ films by in-situ post-growth annealing in the oxygen-containing residual background chamber pressure of ~$10^{-7}$ Torr, as reported in our earlier study.[17] The film's chemical composition was measured using Rutherford backscattering spectroscopy. The oxygen content after annealing was found to scale proportionately with the extent of Mg-excess in the as-deposited film.[17] As-deposited films were produced with a stoichiometry that varies from $Mg_1B_2$ to $Mg_{1.2}B_2$ as a result of differences in the distances of substrates from the Mg Knudsen cell and B electron-beam evaporation source. This method allows us to compare films produced under virtually identical conditions which differ in their oxygen content. The oxygen-containing $MgB_2$ films were coated ex-situ with 1000 Å thick PECVD (Plasma-enhanced chemical vapor deposition) grown $SiO_2$ layer to minimize degradation of $MgB_2$ films and were then photolithographically patterned to a 20 µm wide by 0.5 mm long bridge structure. Measurements of $\rho(T)$ and $\rho(H)$ were made using a 9 T Quantum Design Physical Property Measurement System (PPMS). $H_{c2}$ was defined as $\rho(H_{c2}) = 0.9\rho_n$, and $J_c$ as the current at which the voltage reaches a value corresponding to 1 mV/cm along the film.

Atomic force microscopy (AFM) measurements indicate an increase in grain size by ~50% (55 nm to 75 nm) in high oxygen (>6.8%) films.

$T_c$ is found to decrease linearly with increasing oxygen concentrations up to ~7% (Fig. 1(a)) and then levels off. The observed $T_c$ suppression with increasing oxygen content, therefore, suggests that a fraction of the oxygen is incorporated into the $MgB_2$ grains, most likely as MgO[33] or Mg(B,O)[19] and possibly as substitutional oxygen,[33] and the $T_c$ suppression results from enhanced scattering. If a significant fraction of oxygen is substituted on the boron site, the effect of doping could cause both scattering and a decrease in the density of states at the Fermi level, $N_{Ef}$, with a corresponding drop in $T_c$.[33]

Fig. 1(b) shows the variation in $\Delta\rho$ ($\rho_{300K}$-$\rho_{40K}$) with oxygen concentration in the films. $\Delta\rho$, as pointed out by Rowell,[34] can be used to infer the inter-grain connectivity in $MgB_2$ samples. A linear increase in $\Delta\rho$ has been observed with an increase in oxygen concentration up to ~ 7% in these films indicating that some of the oxygen ends up between the grains as insulating phases such as MgO or boron oxides and resulting in a decreased cross-section of the film carrying current.

The residual resistivity [$\rho_0=\rho(40K)$] can be corrected using Rowell analysis[34] to obtain the intra-grain resistivity values by eliminating the influence of the inter-grain connectivity. Fig. 1(c) shows the monotonic increase of corrected residual resistivity ($\rho_{0,corrected}$) with up to ~7% oxygen concentration resulting possibly from increased carrier scattering from oxygen-containing defects. The observed values in the range of 13-26 $\mu\Omega.cm$ are significantly higher than clean $MgB_2$ (<1 $\mu\Omega.cm$) and are similar in magnitude to some defect-laden films, including ion damaged films, that attain high $H_{c2}$s.[9,10,17,35] The dependence of $T_c$ on the $\rho_{o,corrected}$ is very similar to films damaged progressively by ion irradiation reported in Ref. 35. This further supports our conclusion that the $T_c$ suppression observed in this oxygen doping study can be attributed to scattering by intragranular defects.

The corrected resistivity, as well as Tc (Fig. 1(a)), $\Delta\rho$ (Fig. 1(b)) and $\rho_{0,corrected}$ (Fig. 1(c)), does not change significantly beyond an oxygen concentration of ~7%. The reason for this is not clear, although it may result from the additional oxygen in these films coarsening existing particles rather than introducing new defects.

For comparison to the in-situ annealed films, one Mg-rich $MgB_2$ film was annealed ex-situ in air. This film shows a marginal change in $T_c$ (-10%) (Fig. 1(a)) and a corrected residual resistivity (+10%) (Fig. 1(c)) over vacuum annealed films. $\Delta\rho$, on the other hand, increases to 95 $\mu\Omega.cm$ which is 3 times higher than the in-situ vacuum annealed film with similar oxygen content (Fig. 1(b)) and indicates that only ~8% of the cross-section is carrying current.

$J_c$ is not affected significantly by excess oxygen (>0.65%) and is ~1.5 x $10^6$ A/cm$^2$ at 0 T and 4.2 K in all the samples (Fig. 1(d)). This could be due to the fact that any enhancement in $J_c$ as a result of an increased number of effective pinning centers created by oxygen is compensated by a reduction in $J_c$ due to decreased connectivity and increased grain size[30] in oxygen alloyed films.

Figures 2(a) and 2(b) show the field dependence of the films' $J_c$s at various temperatures. $J_c$s are high (~ $4 \times 10^5$ A/cm$^2$ for both films at 8 T oriented parallel to the film and 4.2 K), even in the presence of large magnetic fields. In all the films, a slight $J_c$ maximum at 0 T is observed at 10 K. We do not understand the mechanism for this.

In large magnetic fields (>2 T), the $J_c$s obtained in our films are much higher than are obtained in alloyed bulk and film MgB$_2$ samples reported in the literature and in NbTi superconducting wires ($3.15 \times 10^5$ A/cm$^2$ at 5 T and $1.15 \times 10^5$ A/cm$^2$ at 8 T),[36] its current rival for magnet applications. Fig. 2(c) compares $J_c$ values obtained in this study with those obtained in bulk samples alloyed with carbon,[14] SiC,[16] and carbon nanotubes[37] and in films alloyed with carbon,[38] oxygen[29] and in clean films grown at low temperatures (270°C).[39] It is interesting to note that our relatively clean film (0.65% oxygen) has much higher $J_c$ at high fields (>4 T) than all other alloyed films and bulk material. This indicates that low growth temperature, which can alter grain structure, plays an important role in improving $J_c$ of MgB$_2$. High $J_c$ obtained by Kitaguchi et al.[39] in their clean films grown at 270°C (Fig. 2(c)) supports this argument.

Fig. 3 shows the flux pinning force ($F_p$) versus applied field (H) plots for MgB$_2$ films. The flux pinning force values increase with deceasing temperatures (Fig. 3(a) and 3(b)). We do not see a clear correlation between pinning force and oxygen concentration in these films (Fig. 3(c)). This may result from a combination of pinning mechanisms, including core-type and $\Delta\kappa$-type. The highest pinning force of 48 GN/m$^3$ was observed in MgB$_2$ film with 2% oxygen at 7.5 T (Fig. 3(c)). This value is 3 times higher than that of NbTi superconductors at 4.2 K.[40]

$dH_{c2}/dT$ (close to $T_c$), apart from limited scatter, increases with increasing oxygen concentration (Fig. 1(e)). Again, as seen earlier for $T_c$, $\Delta\rho$, and $\rho_{0,corrected}$, the rate of change is reduced significantly after oxygen concentrations of ~7 %. $dH_{c2}/dT$ value as high as 1.4 T/K is obtained for the film with 16.5 at.% oxygen in parallel direction (H∥ab). This value is close to $dH_{c2}^{\parallel}/dT$ near $T_c$ of 1.6-1.7 T/K obtained for C-doped MgB$_2$ films with remarkably high $H_{c2}^{\parallel}(0)$ values of ~65 T.[9] Linear extrapolation to 0 K of the $H_{c2}$ vs. T dependence, obtained for the O-alloyed films, yields $H_{c2}^{\parallel}(0)$ values as high as 46 T, while still maintaining reasonably high $J_c$ values (~ $3 \times 10^5$ A/cm$^2$ at 8 T and 10 K). The trend of $dH_{c2}/dT$ dependence on oxygen (Fig. 1(e)) indicates a more pronounced effect of oxygen on $H_{c2}$ when the field is parallel to ab plane compared to when it is perpendicular. These $H_{c2}(0)$ values obtained for O-incorporated (>5% oxygen) $MgB_2$ films, along with earlier reported $H_{c2}(0)$ of cold-grown/annealed $MgB_2$ films[41] and C-doped $MgB_2$ films[9] are quite unusual..

To summarize, oxygen was incorporated in $MgB_2$ films *in situ* by vacuum annealing Mg-rich films. The $T_c$, $\Delta\rho$ and $\rho_{0,corrected}$ dependence with oxygen concentration indicate that the incorporated oxygen is distributed both within and between the grains. $H_{c2}(0)$ and $dH_{c2}/dT$ increase with oxygen content, particularly when the field is applied parallel to the film surface. In contrast, we observe high $J_c$ values which do not change significantly with oxygen content, presumably because this property is dominated by the ingrown disorder present when films are deposited at low growth temperatures (300°C). Our study demonstrates that a combination of oxygen doping and low growth temperature can attain high $J_c$s, in field (~ $3 \times 10^5$ $A/cm^2$ at 8 T and 10 K) and $H_{c2}$ (>45 T), that are needed for next-generation high magnetic field applications.

This work was supported by NSF Grant No. DMR-0514592 and ONR Contracts N00014-05-1-0105 and N00014-06-1-1153. The authors acknowledge use of facilities in the ASU Center for Solid State Science. The authors would like to thank Anil Indluru and Zina Alam for help in AFM studies.

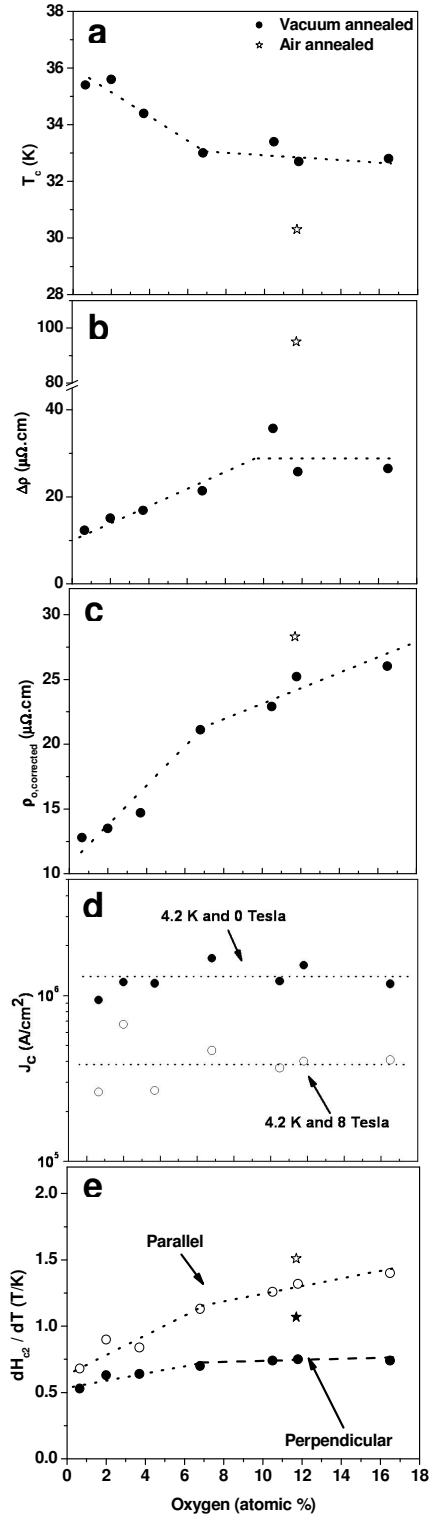

FIG.1 Dependence of (a) $T_c$ (b) $\Delta\rho$ (c) $\rho_{0,corrected}$ (d) $J_c$ and (e) $dH_{c2}/dT$, close to $T_c$, with oxygen concentration in MgB$_2$ films. Dashed lines are intended to serve as guides to the eye, and are not a result of analytical fits to the data.

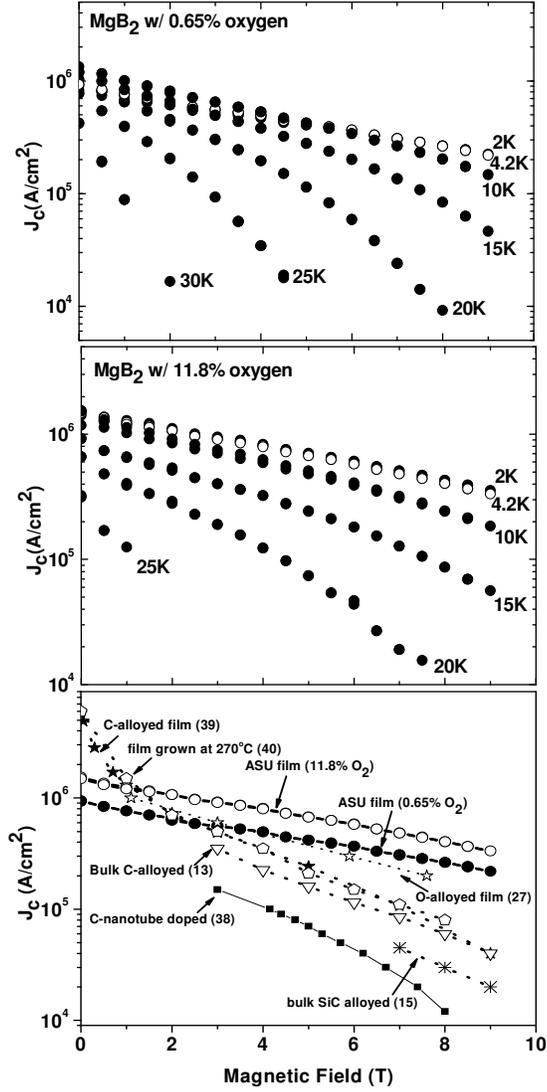

FIG.2 $J_c$ vs. H dependence obtained for $MgB_2$ films with (a) 0.65 at. % oxygen and (b) 11.8 at. % oxygen, at different temperatures, (c) Comparison of $J_c$s obtained in oxygen alloyed films in this study with bulk samples alloyed with carbon,[14] SiC,[16] and carbon nanotubes[37] and films alloyed with carbon,[38] oxygen[29] and grown at low temperature (270°C).[39]

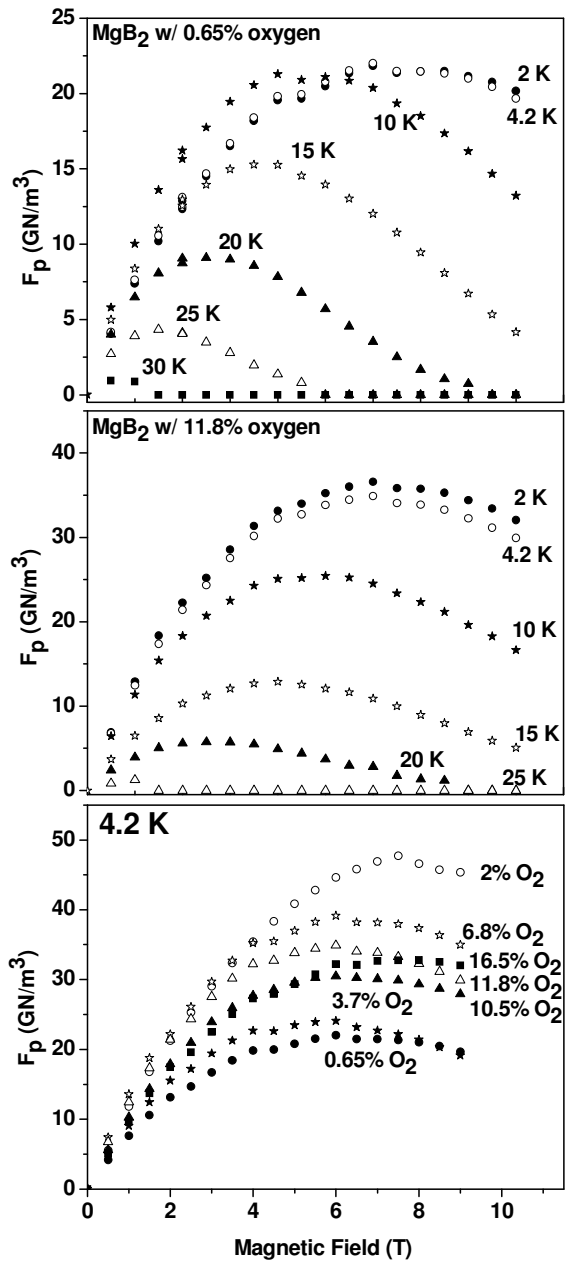

FIG.3 Flux pinning force of MgB$_2$ films with (a) 0.65% and (b) 11.8% oxygen at various temperatures, and (c) with oxygen varying from 0.65% to 16.5% at 4.2 K.